\documentclass[aps,twocolumn,prx]{revtex4} 

\usepackage{amssymb}
\usepackage{amsmath}
\usepackage{graphicx}
\usepackage{color}
\usepackage{setspace}
\usepackage{url}

\definecolor{Gray}{gray}{0.9}
\definecolor{LightCyan}{rgb}{0.88,1,1}
\begin{document} 



\title{\LARGE {\bf A RESTful API for exchanging Materials Data \\in the AFLOWLIB.org consortium}}

\author{Richard H. Taylor$^{1,2}$, Frisco Rose$^2$, Cormac Toher$^2$, Ohad Levy$^{2,\dagger}$, \\
  Marco Buongiorno Nardelli$^3$, Stefano Curtarolo$^{4,\star}$}
\address{$^{1}$National Institute of Standards and Technology, Gaithersburg, Maryland, 20878, USA}
\address{$^{2}$Department of Mechanical Engineering and Materials Science, Duke University, Durham, North Carolina 27708, USA}
\address{$^{3}$Department of Physics and Department of Chemistry, University of North Texas, Denton TX }
\address{$^{4}$Materials Science, Electrical Engineering, Physics and
  Chemistry, Duke University, Durham NC, 27708}
\address{$^{\dagger}$On leave from the Physics department, NRCN, Israel}
\address{$^{\star}${\bf corresponding:} stefano@duke.edu}
\begin{abstract}
  The continued advancement of science depends on shared and reproducible data. In the field of
  computational materials science and rational materials design this entails the construction of
  large open databases of materials properties.
  To this end,
  an \underline{A}pplication \underline{P}rogram \underline{I}nterface ({\small API})
  following REST principles is introduced for the {\small AFLOWLIB}.org materials data repositories consortium.
  {\small AUID}s (\underline{A}flowlib \underline{U}nique \underline{ID}entifier) and
  {\small AURL}s (\underline{A}flowlib \underline{U}niform \underline{R}esource \underline{L}ocator)
  are assigned to the database resources according to a well-defined protocol described herein,
  which enables the client to access, through appropriate queries, the desired data for post-processing.
  This introduces a new level of openness into the {\small AFLOWLIB} repository, allowing the community
  to construct high-level work-flows and tools exploiting its rich data set of calculated structural, thermodynamic,
  and electronic properties.
  Furthermore, federating these tools will open the door to collaborative investigations of unprecedented scope
  that will dramatically accelerate the advancement of computational
  materials design and development. 
\end{abstract}

\date{\today} 
\maketitle 


\section{Introduction}
\label{introduction}
Data-driven materials science has gained considerable traction over the last decade or so.
This is due to the confluence of three key factors:
1) Improved computational methods and tools;
2) Greater computational power; and
3) Heightened awareness of the power of extensive databases in science \cite{nmatHT_editorial}.
The recent Materials Genome Initiative ({\small MGI}) \cite{MGI_OSTP,nmatHT_editorial} reflects the
recognition that many important social and economic
challenges of the 21st century could be solved or mitigated by advanced materials. Computational materials science
currently presents the most promising path to the resolution of these challenges.

The first and second factors above are epitomized by \emph{high-throughput} computation of materials properties
by \emph{ab initio} methods, which is the foundation of an effective approach to materials design and
discovery \cite{nmatHT,Ceder_ScientificAmerican_2013,Greeley2006,curtarolo:TIs,monsterPGM,Fornari_Piezoelectrics_PRB2011,aflowSCINT,aflowTHERMO,YuZunger2012_PRL,curtarolo:art86}.
Recently, the software used to {both manage} the calculation work-flow and perform the analyses have trended toward more public and user-friendly frameworks.
The emphasis is increasingly on portability and sharing of tools and data \cite{aflowPAPER,aflowlibPAPER,materialsproject.org}.
{Similar to the effort presented here, the {\it Materials-Project}  \cite{APL_Mater_Jain2013} 
has been providing open access to its database of computed materials properties through a RESTful API and
a {\it python} library enabling ad-hoc applications \cite{CMS_Ong2012b}.}
{Other examples of online databases for materials properties
  include that being implemented by the Engineering Virtual
  Organization for Cyber Design (EVOCD) \cite{msstate_evocd_website},
  which contains a repository of experimental data, materials
  constants and computational tools for use in Integrated Computational Material Engineering (ICME)}.
The future advance of computational materials science would rely on
interoperable and federatable tools and databases as much as on the
quantities and types of data being produced.

A principle of high-throughput materials science is that one does not know \emph{a priori} where the value of
the data lies for any specific application. Trends and insights are
deduced \emph{a posteriori}. This requires efficient interfaces to
interrogate available data on various levels.
We have developed a simple {\small WEB}-based {\small API} to greatly improve the accessibility and utility of
the {\small AFLOWLIB} database \cite{aflowlibPAPER} to the scientific community.
Through it, the client can access calculated physical properties (thermodynamic, crystallographic, or mechanical properties),
as well as simulation provenance and runtime properties of the included systems.
The data may be used directly (e.g., to browse a class of materials with a desired property) or integrated into higher level work-flows.
The interface also allows for the sharing of updates of data used in previous published works, e.g., previously calculated
alloy phase diagrams
\cite{monster,    
curtarolo:art19,  
curtarolo:art49,
curtarolo:art51,
curtarolo:art57,
curtarolo:art50,
curtarolo:art53,
curtarolo:art54,
curtarolo:art55,
curtarolo:art56,
curtarolo:art62,
curtarolo:art63,
curtarolo:art67}, 
thus the database can be expanded systematically.

\begin{figure*}[t!]
\label{fig1}
\centering
\includegraphics[width=0.999\textwidth]{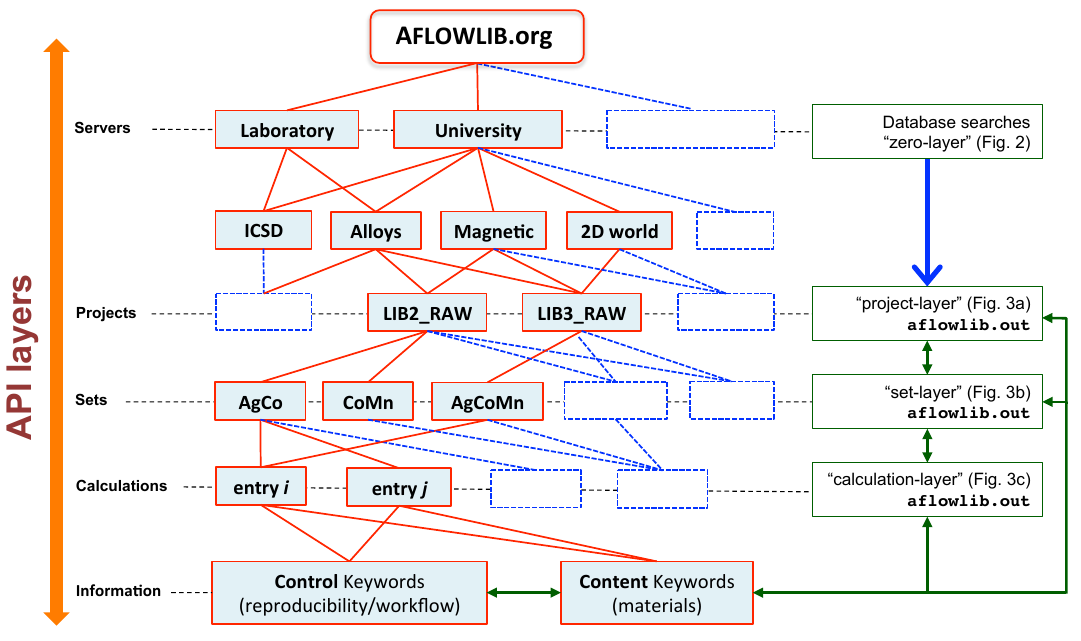}
\caption{\small
 Schematic structure of the {\small AFLOWLIB} consortium}
\end{figure*}

The rest of this paper is organized as follows.
The {\small AFLOWLIB} libraries are presented in Section \ref{AFLOWlibraries}.
A new materials identifier constructed to navigate the libraries is introduced in Section \ref{materials_identifier}.
The data provenance and access format schemes are explained in Sections \ref{data_provenance} and \ref{data_formatting}.
The access syntax and its various options are described in Section
\ref{table_properties}. A few examples for the use of the {\small API} and two computer scripts are given in Section \ref{examples}.
The strategy for updates is mentioned in Section \ref{updates}.
A brief conclusion is included in Section \ref{conclusions}.

\section{The {\small AFLOWLIB} libraries multi-layered structure}
\label{AFLOWlibraries}

At its core, {\small AFLOWLIB} consists of a coordinated set of libraries of first-principles data describing thermodynamic,
structural, and other materials properties of alloy systems (Figure 1).
From the top, it is administered through a large SQL database \cite{SQL} organized in {\it layers}
(reminiscent of other more developed computer interfaces, i.e.,
IEEE-POSIX \cite{POSIX}). Each layer consists of searchable entries
called \verb|aflowlib.out|.
The SQL interface is called the {\it zero-layer} (Figure 2) because of its structural immanence.

In this {\it layered} organization, each \verb|aflowlib.out| can be the child of an \verb|aflowlib.out|-parent or the
parent of an
\verb|aflowlib.out|-child. Each \verb|aflowlib.out| is identified by a name and an address.
The first is the {\small AUID} (\underline{A}flowlib \underline{U}nique \underline{ID}entifier -- \verb|$auid|), while
the second is the {\small AURL} (\underline{A}flowlib \underline{U}niform \underline{R}esource \underline{L}ocator --
\verb|$aurl|). The structure is summarized in Figure 1.

The current implementation of {\small AFLOWLIB} includes three
layers that can be navigated using control keywords and absolute
paths.

\begin{widetext}
\begin{itemize}

\item
{\it Project-layer.} This layer contains information about the project to which the data belongs.
For example, for searches in alloys, the project's {\small AURL} could be of the type \verb|server:AFLOWDATA/| followed by
\url{LIB1_RAW}, \url{LIB1_LIB}, \url{LIB2_RAW}, \url{LIB2_LIB}, \url{LIB3_RAW}, or \url{LIB3_LIB}
corresponding to the single element, binary and
ternary libraries of post- or pre-processed data respectively. For HTTP access \cite{HTML}, the {\small AURL} would be translated from
\verb|$aurl=server:/directory/| into \verb|$web=http://server/directory/|.
Other translations are considered for future implementations.
An example of \verb|aflowlib.out| for the {\it project-layer} is depicted in
Figure 3(a) (with \verb|server=aflowlib.duke.edu|).

\item
{\it Set-layer.} This layer contains information about one or more systems calculated in one or more different configurations
(e.g., various structural prototypes, different unit cells required for phonons calculations using the finite difference method, etc.).
An example of \verb|aflowlib.out| for the {\it set-layer} is depicted in Figure 3(b).
To facilitate reproducibility, species making up the \verb|$aurl| might include a subscript or postfix indicating the
pseudopotential type. For  example, in searches in this layer where the user is interested in the Ag-Ti system, the
{\small AURL} could be of the type
\url{$aurl=server:AFLOWDATA/LIB2_RAW/AgTi_sv/}.
Here the ``\verb|_sv|'' in \verb|Ti_sv| indicates the ``sv'' type of pseudopotential in the quantum code used for the
calculation, in this case {\small VASP} \cite{vasp}.
Other identifiers may be used to indicate pseudopotentials used in, for example, Quantum Espresso ({\small QE}) \cite{qe}
or for potentials coming from other ultrasoft pseudopotential libraries \cite{gbrv}.

\item
{\it Calculation-layer.} This layer contains information about one system calculated in one particular configuration
(e.g.\ AgTi in configuration \verb|$prototype=66| (C11$_b$ \cite{Massalski,navy_crystal_prototypes})).
For entries in this layer, the {\small AURL} could be of the type
\url{$aurl=server:AFLOWDATA/LIB2_RAW/AgTi_sv/66/}.
This \verb|$aurl| points to an \verb|aflowlib.out| describing this specific structure properties.
An example of \verb|aflowlib.out| for the {\it calculation-layer} is depicted in Figure 3(c).
The calculated geometry (here ``\verb|66|'') is one of the strings comprising the {\small AFLOW} prototypes' database.
The prototypes' database acts as a look-up table to common {\it Strukturbericht}
designations \cite{Massalski,navy_crystal_prototypes} and may also refer to geometries included in
the Inorganic Crystal Structure Database (ICSD) \cite{ICSD0,ICSD1,ICSD2}.
In the latter case, a legitimate entry would be something like \verb|.../AgTi_sv/ICSD_58369.AB/|,
where the structure composed of Ag and Ti is calculated in the configuration defined by \verb|$prototype=ICSD_58369|, and the
postfix \verb|.AB| indicates the species ordering: e.g. \verb|.AB| or \verb|.BA| for Ag and Ti in \verb|A|\verb|B| or \verb|B|\verb|A| positions, respectively.
For ternaries the postfix is a combination of \verb|A,B,C|, where indices can be repeated to create binaries (e.g.\ \verb|AAB|). Furthermore, the non-element \verb|X|
can be included to indicate vacancies in sublattices (e.g. from Heusler to half-Heusler systems \cite{curtarolo:art84}).
Valid \verb|$prototype| choices also include strings of the type ``\verb|f123|'' following the enumeration scheme of
Hart and Forcade \cite{enum1,enum2}. Here the characters \verb|f|, \verb|b|, \verb|h|, \verb|s|
indicate \underline{f}cc, \underline{b}cc, \underline{h}cp, and \underline{s}imple cubic structures respectively.
The number that follows designates the enumerated prototype.
The complete list of structure designations can be accessed with the command ``{\sf aflow -{}-protos}'' or by consulting the online links.
The options are illustrated in the {\small AFLOW} manual \cite{aflowPAPER}.
\end{itemize}
\end{widetext}

It is important to note that while the operations of concatenating \verb|strings| to \verb|$aurl| are reminiscent of a UNIX computer file-system, the final \verb|$aurl| is not necessarily served as a directory or a file.
In fact, a {\it computer daemon} (process running in background)
dynamically serves multiple file-system directories for the same HTTP-translated \verb|$aurl|.
The implementation was so chosen in order to better use the {\it calculation-layer} data belonging to different {\it project-layers}
and to allow an \verb|$aurl| to contain multiple servers: e.g. \url{$aurl=server1,server2:somewhere1/AgTi_sv/,somewhere2/AgTi_sv/}.
In future updates of the {\small AFLOWLIB}.org {\small API}, the latter collection will allow users
to download AgTi's data from \verb|server1| and \verb|server2|
concurrently and transparently.

\begin{figure*}[t!]
 \label{fig2}
 \centering
 \includegraphics[width=0.999\textwidth]{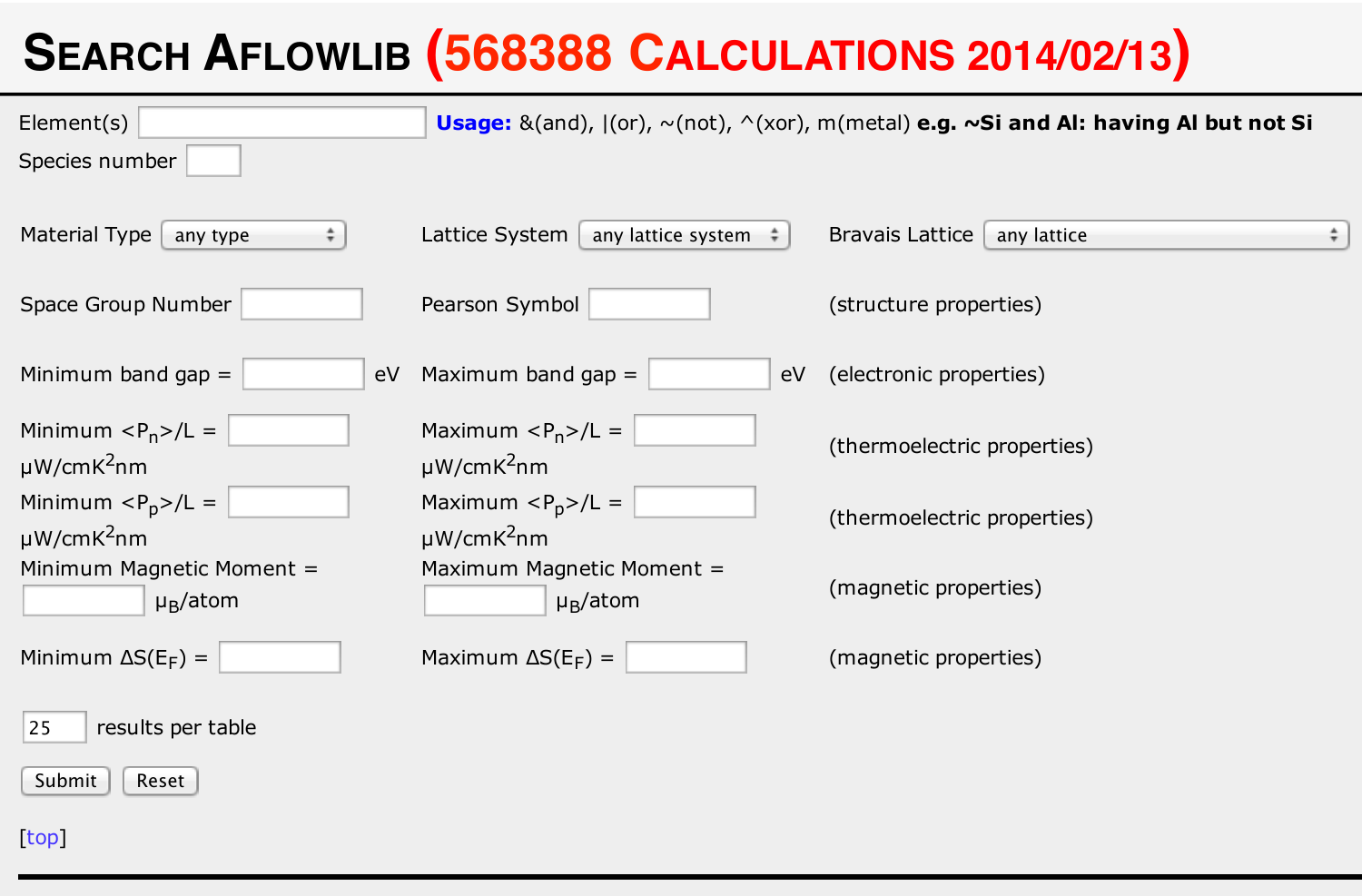}
 \caption{\small Project {\small WEB}-interface search form: {\it zero-layer}.}
\end{figure*}

\begin{figure*}[t!]
 \label{fig3}
 \centering
 \includegraphics[width=1.0\textwidth]{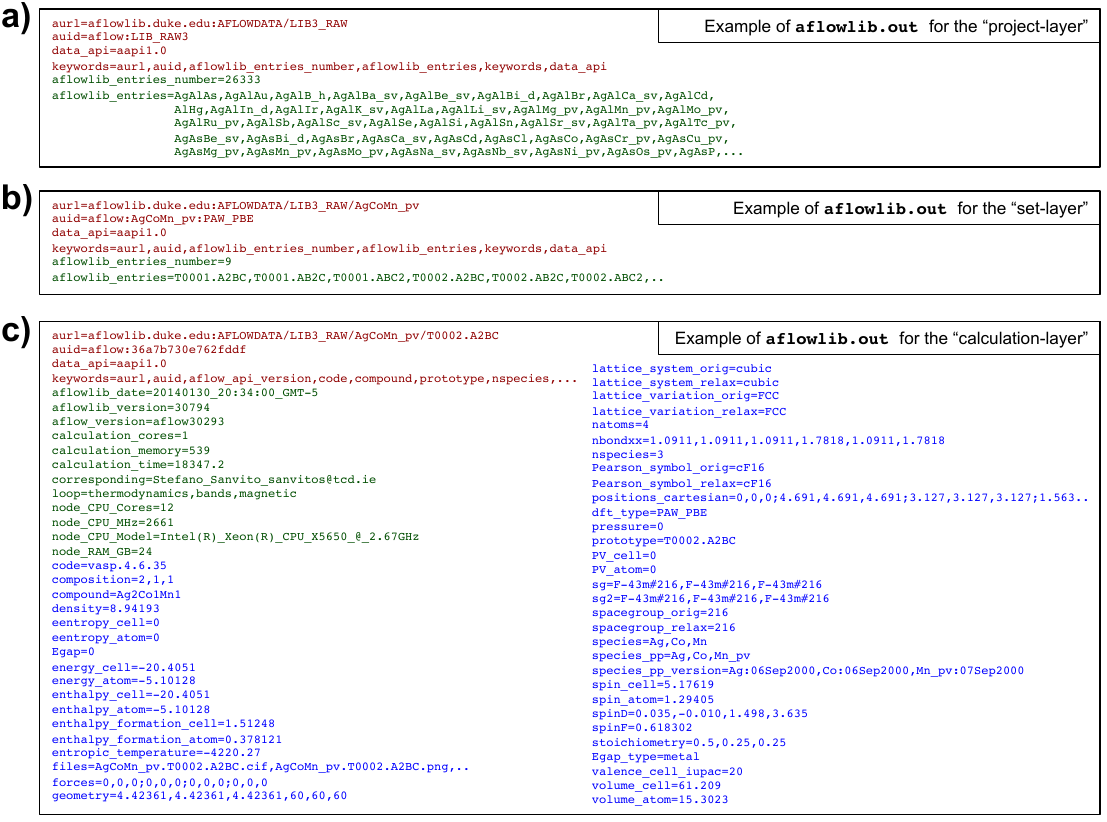}
 \caption{\small
  Examples of the contents defining ``aflowlib.out" entries
  for the {\it project-layer} (panel {\bf a}),  for the {\it set-layer} (panel {\bf b}),
  and for the {\it calculation-layer} (panel {\bf c}). 
  As per {\small AFLOWLIB REST-API} version 1.0 (this document), mandatory keywords are listed in red, optional control keywords are in green, and optional materials keywords are in blue.}
\end{figure*}

\section{Materials Object Identifier: \underline{A}flowlib \underline{U}nique \underline{ID}entifier}
\label{materials_identifier}
Following the spirit of the Digital Object Identifier (DOI) \cite{DOI} every entry in our database is identified by an alpha-numeric
string called an {\small AUID} (\underline{A}flowlib \underline{U}nique \underline{ID}entifier -- \verb|$auid|).
In future publications, we will use the {\small AUID} as a permanent string to the correct {\small AURL}, which might get relocated across servers.
By giving a short {\small AUID}, the user can identify
1) {\small WEB} locations for the deliverables,
2) appropriate points of contact (name, emails or other means of identifying corresponding authors), and
3) copyright and publication date.
We have chosen this linked approach so that future databases and expansions of the current ones
(e.g., locations, servers, personnel, etc.) will not affect the retrieval of the objects.
A simple {\small WEB-FORM} will be offered to the community to insert {\small AUID}s and to retrieve the information.
The {\small AUID} is constructed from a 64-bit CRC checksum (cyclic redundancy check \cite{CRC61})
of concatenated input and output files.
For example, the bcc Ag-Ti structure \verb|66| described above would be accessed
using the {\small AUID} ``\url{$auid=aflow:448e178e19e9e973}''.
The negligible probability of duplicates with this checksum length,
and the opportunity of adding a ``void character" to the input files (a grain of salt)
guarantee the uniqueness of the {\small AUID} and its expandibility for many years to come.
We welcome other groups to set up versions of {\small AUID}s and to communicate with us the necessary
identification data so that our consortium can offer {\small WEB} links to entries beyond the ones prepared by
its current members.

\section{Data Provenance}
\label{data_provenance}

Reproducibility is a key tenet of the scientific method, but replication occurs rarely.
As experiments become more complex and costly this problem will only worsen.
A particular advantage of simulated experiments is that they are uniquely suited for reproducibility.
Imagine a scenario in which the original experimenter publishes his/her
executable workflow with the necessary input data.
Then, anyone wishing to validate or build upon those results does so with a single command on a
system with access to the appropriate software.
This could be done so that all the software tools, input and output data are maintained remotely,
lowering cost, improving ecological sustainability (saving electricity) and increasing collaboration.

In reality, such a framework is not yet available and its implementation is not so straightforward.
Despite the potential advantages of computational science in the realm of reproducibility,
the crucial data provenance is often completely neglected.  
An overall culturally-driven disincentive to the actual reproduction of experiments is pervasive.
Fortunately there are indications that this problem is being recognized and addressed \cite{MGI_OSTP}.

{\small AFLOWLIB} data is reproducibly structured---the workflow and
input parameters are defined by the {\small AFLOW} software
and a single master input file \texttt{aflow.in}, leading to easily reproducible input parameters.
In the case of Density Functional Theory-based simulations, this includes essential calculation parameters
such as the {\bf k}-mesh density, the energy cut-off, the exchange correlation potential, the {\it ab initio} software used,
and the geometry of the structures.
Combining the parameters contained in this file with other provenance related data (calculation time, memory, code, etc.),
the {\small AFLOW-API} provides \emph{curated and reproducible data}.

\section{Data formatting, ownership, and a federatable framework}
\label{data_formatting}

Data access can be obtained at any level through the {\small API} with the appropriate {\small AURL} strings, currently translated into {\small WEB} inquiries.
{\small WEB} forms such as the one shown in Fig 2 allow the user to search within a project for
data fitting specified criteria.
Alternatively, access through the {\small API} is supported by several data formats:
``HTML'' \cite{HTML}, ``JSON'' \cite{JSON}, ``DUMP'', ``PHP'' \cite{PHP}, ``TEXT'', and ``NONE''.

\begin{description}
\item \textbf{``HTML''} (default): \url{$aurl/?format=html}.
\item \textbf{``JSON''} (javascript syntax): \url{$aurl/?format=json}.
\item \textbf{``DUMP''} (php constructs): \url{$aurl/?format=dump}.
\item \textbf{``PHP''} (valid php syntax): \url{$aurl/?format=php}.
\item \textbf{``TEXT''} (no syntax): \url{$aurl/?format=text}.
\item \textbf{``NONE''} (no output): \url{$aurl/?format=none}.
\end{description}
 
The format option is intended for use on the level of the entry returning the whole  \url{aflowlib.out}.
For any given property in the set of keywords contained in the entry, the mode is currently to return a simple
byte sequence with no formatting. Attempting to format a single property returns the full property set in the specified format.
For example, the {\small AURL} \url{$aurl/?density} returns the density of the specified structure.
More specifically, for \url{$aurl=aflowlib.duke.edu:AFLOWDATA/LIB2_RAW/AgTi_sv/66/}\url{?density} it returns ``6.54346''.
However, \url{.../AgTi_sv/66/?density&format=html} is equivalent to \url{.../AgTi_sv/66/?format=html}, returning the
full set of entry properties.
``HTML'' is primarily for interactive use where keywords and files can be promptly explored with web browsers;
``JSON'' and ``PHP'' are valid language syntaxes to facilitate the data access by programmed codes;
``DUMP'' allows the user to access in his/her own method;
``TEXT''  returns the entry in a single line with the keywords separated by ``\url{|}'' so that other databases can be built on top of {\small AFLOWLIB}.org;
and ``NONE'' may be useful as a method to test the existence of an {\small AURL} or for debug purposes.

Clear attribution of contributed data is essential for the
development of distributed databases comprising inputs from a wide
network of contributors. {\small AFLOW} facilitates attribution with
the {\small AUID}, a unique and persistent identifier, that
includes the author, laboratory, group, and affiliation as data
entry fields. The shared content in the database is simple to
reduce or augment according to a contributor's preference and
the attribution is ensured by the unique identifier and
contributor labels that are accessible with the {\small AFLOWLIB}-{\small API}.

The structure of the {\small AFLOWLIB} database is \emph{federated}:  Autonomous members of the consortium (with distinct geographical locations and affiliations)
are able to transparently contribute to a composite database, preserving ownership and claim over the substance of their data. The underlying meta-data schema of the
contributed data are consistent by production, to ensure the clarity and searchability of the composite database.
A contributor to the consortium begins by downloading the latest version of the {\small AFLOW} binary (as of writing this paper this is version 30825) and interfacing with a quantum code.
{\small AFLOW} is currently configured to run {\small VASP} automatically. Pre- and post-processing is functional for both  {\small VASP}  and Quantum Espresso so that
agnostic standardization of inputs and outputs between the two codes can be obtained.

{The layered and {reentrant} structure of the {\small AFLOWLIB} {\small API} allows the {manipulation}
of data coming from different sources and databases, {e.g.} the {\it Materials Project} \cite{APL_Mater_Jain2013,CMS_Ong2012b}.
{In order to facilitate} this future extension, special
keywords are {introduced here} to identify the source (\url{$aurl/?data_source}) and the translated syntax of the information (\url{$aurl/?data_language}).
We foresee a global common interface where users can approach heterogeneous data and applications to leverage the efforts of different consortia.
Note that in this scenario, due diligence is required to recognize the authorship of the original work, and not the serving database, merely.}

\begin{widetext}
\section{Table of Properties and API Keywords}
\label{table_properties}
This section includes the keywords currently present in the database: description, type, inclusion policy and the {\small  AFLOWLIB} syntax for retrieval.
The list is divided into mandatory, optional control and optional materials keywords. The mandatory keywords must be present in every entry at all layers of the database. 
Some of the optional control keywords appear at the {\it projects} and {\it systems} levels while others appear at the {\ calculations} level. 
The optional materials keywords usually appear just at the {\it calculations layer}, and not all of them are present in all of the entries.
Each entry begins with the {\small AURL} and {\small AUID} keywords, denoted by the syntax words \verb|$aurl| and  \verb|$auid|, respectively.

\def\description{\item {{\it Description.}\ }}
\def\type{\item {{\it Type.}\ }}
\def\example{\item {{\it Example.}\ }}
\def\inclusionmandatory{\item {{\it Inclusion.} \url{mandatory}}}
\def\inclusionoptional{\item {{\it Inclusion.} \url{optional}}}
\def\units{\item {{\it Units.}\ }}
\def\tol{\item {{\it Tolerance.}\ }}
\def\syntax{\item {{\it Request syntax.}\ }}

\def\ENERGYunit{e.g., eV or Ry if the calculations were performed with {\small VASP} \cite{vasp} or {\small QE} \cite{qe}, respectively}
\def\ENERGYunitatom{e.g., eV or Ry (eV/atom or Ry/atom) if the calculations were performed with {\small VASP} \cite{vasp} or {\small QE} \cite{qe}, respectively}
\def\LLLunits{e.g., \AA$^3$ or Bohr$^3$ (\AA$^3$/atom or Bohr$^3$/atom) if the calculations were performed with {\small VASP} \cite{vasp} or {\small QE} \cite{qe}, respectively}
\def\FORCEunit{e.g., eV/\AA\ or a.u if the calculations were performed with {\small VASP} \cite{vasp} or {\small QE} \cite{qe}, respectively}
\def\Lunits{e.g., \AA\ or a.u. (Bohr) if the calculations were performed with {\small VASP} \cite{vasp} or {\small QE} \cite{qe}, respectively}
\def\Punits{e.g., kbar or a.u. (Ry/Bohr) if the calculations were performed with {\small VASP} \cite{vasp} or {\small QE} \cite{qe}, respectively}
\def\Sunits{e.g., $\mu_B$ (Bohr magneton)}

\subsection{Mandatory keywords}
\label{mandatory_keywords}

\begin{itemize}
  
\item
  \verb|/| or \url{/?format=html}, \url{json}, \url{dump}, \url{php}, or \url{text}.
  \begin{itemize}
    \description The whole entry. It can be read in different formats. For the {\it calculation-layer} it can also be read as \url{/?aflowlib.out}.
    \type lines of \verb|strings|.
    \example  See examples of entries at different levels in Figure 3.
    \syntax \verb|$aurl|.
  \end{itemize}
  
\item
  \verb|auid|
  \begin{itemize}
    \description ``AFLOWLIB Unique Identifier'' for the entry, {\small AUID}, which can be used as a publishable object identifier,
    following the spirit of the DOI foundation \cite{DOI} (see Section \ref{materials_identifier}).
    \type \verb|string|.
    \example \verb|auid=aflow:e9c6d914c4b8d9ca|.
    \syntax \verb|$aurl/?auid|.
  \end{itemize}
  
\item
  \verb|aurl|
  \begin{itemize}
    \description ``AFLOWLIB Uniform Resource Locator'' returns the {\small AURL} of the entry. The web server is separated from the web directory with ``:''. 
    This tautological keyword, \verb|aurl| returning itself, is useful for debug and hyperlinking purposes.
    \type \verb|string|.
    \example \verb|aurl=aflowlib.duke.edu:AFLOWDATA/LIB3_RAW/Bi_dRh_pvTi_sv/T0003.ABC:LDAU2|.
    \syntax \verb|$aurl/?aurl|.
  \end{itemize}
  
\item
  \verb|data_api|
  \begin{itemize}
    \description ``AFLOWLIB'' version of the entry, {\small API}. This article describes version 1.0 of the {\small REST-API}.
    \type \verb|string|.
    \example \verb|data_api=aapi1.0|.
    \syntax \verb|$aurl/?data_api|.
  \end{itemize}
  
\item
  \verb|keywords|
  \begin{itemize}
    \description This includes the list of keywords available in the entry, separated by commas. All of the keywords can be requested to the database.
    The request \verb|keywords| should be the first one made, so that the reader is made aware of the available keywords.
    \type List of \verb|strings| separated by ``,''.
    \example \verb|keywords=aurl,auid,loop,code,compound,prototype,nspecies,natoms,...|.
    \syntax \verb|$aurl/?keywords|.
  \end{itemize}
  
\end{itemize}

\subsection{Optional controls keywords (alphabetic order)}
\label{optional_keywords_control}

\begin{itemize}

\item
 \verb|aflowlib_entries| (\verb|aflowlib_entries_number|)
\begin{itemize}
 \description For {\it projects} and {\it set-layer} entries (see Figure 1), \verb|aflowlib_entries| lists the available sub-entries which are associated with the \verb|$aurl| of the subdirectories.
 By parsing \verb|$aurl/?aflowlib_entries| (containing \verb|$aurl/aflowlib_entries_number| entries) the user finds the further locations to interrogate.
 \type Set of \verb|strings| separated by ``,'' (\verb|number|).
 \example \verb|aflowlib_entries=AgAl,AgAs,AgAu,AgB_h,AgBa_sv,AgBe_sv,AgBi_d,AgBr,AgCa_sv,...| (\verb|aflowlib_entries_number=1524|).
 \syntax \verb|$aurl/?aflowlib_entries| (\verb|$aurl/aflowlib_entries_number|).
\end{itemize}

\item
\verb|aflowlib_date| (\verb|aflowlib_version|)
\begin{itemize}
\description Returns the date (version) of the {\small AFLOW} post-processor which generated the entry for the library. This entry is useful for debugging and regression purposes.
\type \verb|string|.
\example \verb|aflowlib_date=20140204_13:10:39_GMT-5| (\verb|aflowlib_version=30794|).
\syntax \verb|$aurl/?aflowlib_date| (\verb|$aurl/?aflowlib_version|).
\end{itemize}

\item
\verb|aflow_version|
\begin{itemize}
\description Returns the version number of {\small AFLOW} used to perform the calculation. This entry is useful for debugging and regression purposes.
\type \verb|string|.
\example \verb|aflow_version=aflow30641|.
\syntax \verb|$aurl/?aflow_version|.
\end{itemize}

\item
\verb|author|
\begin{itemize}
\description Returns the name (not necessarily an individual) and affiliation associated with authorship of the data.
Multiple entries are separated by commas. Spaces are substituted with ``\_'' to aid parsing.
\type List of \verb|strings| separated by ``,''.
\example \verb|author=Marco_Buongiorno_Nardelli,Ohad_Levy,Jesus_Carrete|.
\syntax \verb|$aurl/?author|.
\end{itemize}

\item
\verb|calculation_cores|, \verb|calculation_memory|, \verb|calculation_time|
\begin{itemize}
\description Number of processors/cores, maximum memory, total time used for the calculation.
\type \verb|number|, \verb|number|, \verb|number|.
\units adimensional, Megabytes, seconds.
\example \verb|calculation_cores=32|, \verb|calculation_memory=8376.13|, \verb|calculation_time=140713|.
\syntax \verb|$aurl/?calculation_cores|, \verb|$aurl/?calculation_memory|, \verb|$aurl/?calculation_time|.
\end{itemize}

\item
\verb|corresponding|
\begin{itemize}
\description Returns the name (not necessarily an individual) and affiliation associated with the data origin concerning correspondence about data.
Multiple entries are separated by commas. Spaces are substituted with ``\_'' to aid parsing.
\type List of \verb|strings| separated by ``,''.
\example \verb|corresponding=M_Buongiorno_Nardelli_mbn@unt.edu|.
\syntax \verb|$aurl/?corresponding|.
\end{itemize}

\item
\verb|data_source|, \verb|data_language|
\begin{itemize}
  \description %
  {As mentioned in the text, the layered structure of {\small AFLOWLIB} well adapts to serve and translate data presented in other open databases.
    If this is the case, the source and language ({\small API}) of the data are given with these two keywords.
    When using non-{\small AFLOWLIB} data, due diligence is required to recognize the authorship of the original work, and not the serving database, merely.
}
\type \verb|strings|.
\example \verb|data_source=aflowlib|  \verb|data_language=translated|
\syntax \verb|$aurl/?data_source| and \verb|$aurl/?data_language|.
\end{itemize}

\item
\verb|loop|
\begin{itemize}
\description Informs the user of the type of post-processing that was performed.
\type List of \verb|strings| separated by ``,''.
\example \verb|loop=thermodynamics,bands,magnetic|.
\syntax \verb|$aurl/?loop|.
\end{itemize}

\item
\verb|node_CPU_Cores,| \verb|node_CPU_MHz|, \verb|node_CPU_Model|, \verb|node_RAM_GB|
\begin{itemize}
\description Information about the node/cluster where the calculation was performed. Number of cores, speed, model, and total memory accessible to the calculation.
\type\verb|number|, \verb|number|, \verb|string|, \verb|number|.
\units MHz for speed, gigabytes for RAM.
\example \verb|node_CPU_Cores=12|,
\verb|node_CPU_MHz=2661|,
\verb|node_RAM_GB=48|, \\
\verb|node_CPU_Model=Intel(R)_Xeon(R)_CPU_X5650_@_2.67GHz|.
\syntax \verb|$aurl/?node_CPU_Cores|, \verb|$aurl/?node_CPU_MHz|, \\ \verb|$aurl/?node_CPU_Model|, \verb|$aurl/?node_RAM_GB|.
\end{itemize}

\item
\verb|sponsor|
\begin{itemize}
\description Returns information about funding agencies and other sponsors for the data.
Multiple entries are separated by commas. Spaces are substituted with ``\_'' to aid parsing.
\type List of \verb|strings| separated by ``,''.
\example \verb|sponsor=DOD_N000141310635,NIST_70NANB12H163|.
\syntax \verb|$aurl/?sponsor|.
\end{itemize}

\end{itemize}

\subsection{Optional materials keywords (alphabetic order)}
\label{optional_keywords_materials}

\begin{itemize}

\item
\verb|Bravais_lattice_orig| (\verb|Bravais_lattice_relax|)
\begin{itemize}
\description Returns the Bravais lattice \cite{tables_crystallography_A} of the original unrelaxed (relaxed) structure before (after) the calculation.
\type \verb|string|.
\example \verb|Bravais_lattice_orig=MCLC| (\verb|Bravais_lattice_relax=MCLC|).
\syntax \verb|$aurl/?Bravais_lattice_orig| (\verb|$aurl/?Bravais_lattice_relax|).
\tol {Calculations of lattices (Brillouin zones), prototypes, and symmetries (point/factor/space groups) are based on 
  different algorithms and require different sets of tolerances.
  To guarantee self-consistency of the results, initial tolerances are set to very stringent values ({\it e.g.}, 10$^{-4}$\% for distances, 10$^{-2}$\% for angles, 10$^{-4}$\% for spectral radii of mapping matrices, etc.) and slowly increased alternatingly (by a factor of 2) until self-consistency is found amongst geometrical descriptors. The final tolerances are usually of the order of $\sim0.5\%$ for distances and $\sim1\%$ for angles. }
\end{itemize}

\item
\verb|code|
\begin{itemize}
\description Returns the software name and version used to perform the simulation.
\type \verb|string|.
\example \verb|code=vasp.4.6.35|.
\syntax \verb|$aurl/?code|.
\end{itemize}

\item
\verb|composition|
\begin{itemize}
\description Returns a comma delimited composition description of the structure entry in
the calculated cell.
\type List of \verb|number| separated by ``,''.
\example \verb|composition=2,6,6|. (For a A$_2$B$_6$C$_6$ compound).
\syntax \verb|$aurl/?composition|.
\end{itemize}

\item
\verb|compound|
\begin{itemize}
\description Similar to \verb|composition|. Returns the composition description of the compound in the calculated cell.
\type Set of $\{$\verb|string|$\cdot$\verb|number|$\}$.
\example \verb|compound=Co2Er6Si6|.
\syntax \verb|$aurl/?compound|.
\end{itemize}

\item
\verb|density|
\begin{itemize}
\description Returns the mass density.
\type \verb|number|.
\units grams/cm$^3$.
\example \verb|density=7.76665|.
\syntax \verb|$aurl/?density|.
\end{itemize}

\item
\verb|dft_type|
\begin{itemize}
\description
Returns information about the pseudopotential type,
the exchange correlation functional used (normal or hybrid) and use of GW.
\type Set of \verb|strings| separated by ``,''.
\example If the calculations were performed with {\small VASP} \cite{vasp}, the entry could include ``US", ``GGA'', ``PAW\_LDA", ``PAW\_GGA", ``PAW\_PBE" , ``GW'', ``HSE06'' (February 2014).
\example \verb|dft_type=PAW_PBE,HSE06|.
\syntax \verb|$aurl/?dft_type|.
\end{itemize}

\item
\verb|eentropy_cell| (\verb|eentropy_atom|)
\begin{itemize}
\description Returns the electronic entropy of the unit cell used to converge the \emph{ab initio} calculation (smearing).
\type \verb|number|.
\units Natural units of the \verb|$code|, \ENERGYunitatom.
\example \verb|eentropy_cell=0.0011| (\verb|eentropy_atom=0.0003|).
\syntax \verb|$aurl/?eentropy_cell| (\verb|$aurl/?eentropy_atom|).
\end{itemize}

\item
\verb|Egap|
\begin{itemize}
\description Band gap calculated with the approximations and pseudopotentials described by other keywords.
\type \verb|number|.
\units eV.
\example \verb|Egap=2.5|.
\syntax \verb|$aurl/?Egap|.
\end{itemize}

\item
\verb|Egap_fit|
\begin{itemize}
\description Simple cross-validated correction (fit) of \verb|Egap|. See Ref. \cite{aflowSCINT} for the definition.
\type \verb|number|.
\units eV.
\example \verb|Egap_fit=3.5|.
\syntax \verb|$aurl/?Egap_fit|.
\end{itemize}

\item
\verb|Egap_type|
\begin{itemize}
\description Given a band gap, this keyword describes if the system is a metal, a semi-metal, an insulator with direct or indirect band gap.
\type \verb|string|.
\example \verb|Egap_type=insulator_direct|.
\syntax \verb|$aurl/?Egap_type|
\end{itemize}

\item
\verb|energy_cell| (\verb|energy_atom|)
\begin{itemize}
\description Returns the total \emph{ab initio} energy of the unit cell $E$ (energy per atom ---the value of \verb|energy_cell|/$N$).
\type \verb|number|.
\units Natural units of the \verb|$code|, \ENERGYunitatom.
\example \verb|energy_cell=-82.1656| (\verb|energy_atom=-5.13535|).
\syntax \verb|$aurl/?energy_cell| (\verb|$aurl/?energy_atom|).
\end{itemize}

\item
\verb|energy_cutoff|
\begin{itemize}
\description Set of energy cut-offs used during the various steps of the calculations.
\type Set of \verb|strings| separated by ``,'';
\units Natural units of the \verb|$code|, \ENERGYunitatom.
\example \verb|energy_cutoff=384.1,384.1,384.1|.
\syntax \verb|$aurl/?energy_cutoff|.
\end{itemize}

\item
\verb|enthalpy_cell| (\verb|enthalpy_atom|)
\begin{itemize}
\description Returns the enthalpy of the system of the unit cell $H = E + PV$ (enthalpy per atom ---the value of \verb|enthalpy_cell|/$N$).
\type \verb|number|.
\units Natural units of the \verb|$code|, \ENERGYunitatom.
\example \verb|enthalpy_cell=-82.1656| (\verb|enthalpy_atom=-5.13535|).
\syntax \verb|$aurl/?enthalpy_cell| (\verb|$aurl/?enthalphy_atom|).
\end{itemize}

\item
\verb|enthalpy_formation_cell| (\verb|enthalpy_formation_atom|)
\begin{itemize}
\description Returns the formation enthalpy $\Delta H_F$ per unit cell ($\Delta {H_F}_{atomic}$ per atom). For compounds $A_{N_A}B_{N_B}\cdots$ with $N_A+N_B\cdots=N$ atoms per cell,
this is defined as: $\Delta H_F \equiv E(A_{N_A}B_{N_B}\cdots)-\left[N_AE(A)_{atomic}+N_BE(B)_{atomic}+\cdots\right]$ (in the \verb|_atom| case with $A_{x_A}B_{x_B}\cdots$ and
$x_A+x_B\cdots=1$ we have $\Delta {H_F}_{atomic} \equiv E(A_{x_A}B_{x_B}\cdots)_{atomic}-\left[x_AE(A)_{atomic}+x_BE(B)_{atomic}+\cdots\right]$).
\type \verb|number|.
\units Natural units of the \verb|$code|, \ENERGYunitatom.
\example \verb|enthalpy_formation_cell=-33.1587| (\verb|enthalpy_formation_atom=-0.720841|).
\syntax \verb|$aurl/?enthalpy_formation_cell| (\verb|$aurl/?enthalpy_formation_atom|).
\end{itemize}

\item
\verb|entropic_temperature|
\begin{itemize}
\description Returns the entropic temperature as defined in Ref. \cite{nmatHT,monsterPGM} for the structure. 
The analysis of formation enthalpy is, by itself, insufficient to compare alloy stability at different concentrations
and their resilience toward high-temperature disorder. The formation enthalpy represents the
ordering-strength of a mixture $A_{x_A}B_{x_B}C_{x_C}\cdots$ against decomposition into its pure constituents at the appropriate concentrations $x_A$, $x_B$ $x_C$, $\cdots$.
($\Delta H_F$ is negative for compound forming systems). However, it does not contain information about its resilience against
disorder, which is captured by the entropy of the system. To quantify this resilience we define the {\it entropic temperature} for each compound as:
\begin{equation}
T_s(A_{x_A}B_{x_B}C_{x_C}\cdots)\equiv\left\{\frac{\Delta H_F(A_{x_A}B_{x_B}C_{x_C}\cdots)}{k_B\left[{x_A} \log ({x_A})+{x_B} \log ({x_B})+{x_C} \log ({x_C})+\cdots\right]}\right\},
\end{equation}
where the sign is chosen so that a positive temperature is needed for competing against compound stability.
This definition assumes an ideal scenario \cite{nmatHT} where the entropy is
$S\left[\left\{{x_i}\right\}\right]=-k_B\sum_i{x_i}\log ({x_i})$.
$T_s$ is a concentration-maximized formation enthalpy weighted by the inverse of its entropic contribution.
Its maximum $T_s={\rm max}_{phases}\left[T_s(phases)\right]$ represents the deviation of a system convex-hull from
the purely entropic free-energy hull, $-TS(x)$, and hence
the ability of its ordered phases to resist the temperature-driven deterioration into a disordered mixture
exclusively promoted by configurational-entropy.
\type \verb|number|.
\units Kelvin.
\example \verb|entropic_temperature=1072.1|.
\syntax \verb|$aurl/?entropic_temperature|.
\end{itemize}

\item
\verb|files|
\begin{itemize}
\description Provides access to the input and output files used in the simulation (provenance data).
\type List of \verb|strings| separated by ``,''.
\example
\verb|files=Bi_dRh_pv.33.cif,Bi_dRh_pv.33.png,CONTCAR.relax,CONTCAR.relax1,|\\
\verb|DOSCAR.static.bz2,EIGENVAL.bands.bz2,KPOINTS.bands.bz2,aflow.in,edata.bands.out,|\\
\verb|edata.orig.out,edata.relax.out,...|
\syntax \verb|$aurl/?files|.
\description Once the ``\verb|files|'' list has been parsed, each file can be accessed with \verb|$aurl/file| (note no ``\verb|?|" for accessing individual files).
\end{itemize}

\item
\verb|forces|
\begin{itemize}
\description Final quantum mechanical forces $(F_i,F_j,F_k)$ in the notation of the code.
\type Triplets (\verb|number|,\verb|number|,\verb|number|) separated by ``;'' for each atom in the unit cell.
\units Natural units of the \verb|$code|, \FORCEunit.
\example
\verb|forces=0,-0.023928,0.000197;0,0.023928,-0.000197;...|
\syntax \verb|$aurl/?forces|. 
\end{itemize}

\item
\verb|geometry|
\begin{itemize}
\description Returns geometrical data describing the unit cell in the usual $a,b,c,\alpha,\beta,\gamma$ notation ($\alpha\equiv\widehat{\{\vec{b},\vec{c}\}}, \beta\equiv\widehat{\{\vec{c},\vec{a}\}}, \gamma\equiv\widehat{\{\vec{a},\vec{b}\}}$.
\type Sixtuplet (\verb|number|,\verb|number|,\verb|number|,\verb|number|,\verb|number|,\verb|number|).
\units $a,b,c$ are the natural units of the \verb|$code|, \Lunits. $\alpha,\beta,\gamma$ are in degrees.
\example \verb|geometry=18.82,18.82,18.82,32.41,32.41,32.41|.
\syntax \verb|$aurl/?geometry|.
\end{itemize}

\item
\verb|lattice_system_orig|, \verb|lattice_variation_orig| \\ (\verb|lattice_system_relax|, \verb|lattice_variation_relax|)
\begin{itemize}
\description Return the lattice system \cite{tables_crystallography_A,bilbao} and lattice variation (Brillouin zone) of the original-unrelaxed (relaxed) structure before (after) the calculation.
See Ref. \cite{aflowBZ,aflowlibPAPER} for the lattice variation and Brillouin zones notations.
\type \verb|string|, \verb|string|.
\example \verb|lattice_system_orig=rhombohedral|, \verb|lattice_variation_orig=RHL1| \\ (\verb|lattice_system_relax=monoclinic|, \verb|lattice_variation_relax=MCLC1|).
\syntax \verb|$aurl/?lattice_system_orig|, \verb|$aurl/?lattice_variation_orig|\\ (\verb|$aurl/?lattice_system_relax|, \verb|$aurl/?lattice_variation_relax|).
\tol {See entry} \verb|Bravais_lattice_orig| {or discussion about tolerances.}
\end{itemize}

\item
\verb|kpoints|
\begin{itemize}
\description Set of {\bf k}-point meshes uniquely identifying the various steps of the calculations,
e.g.\ relaxation, static and electronic band structure (specifiying the {\bf k}-space symmetry points of
the structure).
\type Set of \verb|numbers| and \verb|strings| separated by ``,'' and ``;''.
\example \verb|kpoints=10,10,10;16,16,16;G-X-W-K-G-L-U-W-L-K+U-X|.
\syntax \verb|$aurl/?kpoints|.
\end{itemize}

\item
\verb|ldau_TLUJ|
\begin{itemize}
\description This vector of numbers contains the parameters of the ``DFT+U'' calculations, based on a corrective functional inspired by the Hubbard model \cite{LDAU,reviewLDAU}.
Standard values in the {\small AFLOWLIB}.org library come from Refs. \cite{aflowSCINT,aflowTHERMO}.
There are four fields (\verb|T|;\verb|{L}|;\verb|{U}|;\verb|{J}|), separated by ``;''.
The first field indicates the type (\verb|T|) of the DFT+U corrections:
 type=1, the rotationally invariant version introduced by Liechtenstein {\it et al.} \cite{LDAU1};
type=2, the simplified rotationally invariant version introduced by Dudarev {\it et al.} \cite{LDAU2}.
The second field indicates the $l$-quantum number (\verb|{L}|, one
number for each species separated by ``,'') for which the on-site
interaction is added (-1=neglected,
0=$s$, 1=$p$, 2=$d$, 3=$f$).
The third field lists the effective on-site Coulomb interaction parameters (\verb|{U}|, one number for each species separated by ``,'').
The fourth field specifies the effective on-site exchange interaction parameters (\verb|{J}|, one number for each species separated by ``,'').
Although more compact, the convention is similar to the {\small VASP} notation \cite{vasp}.
\units a-dimensional; \{adimensional\}; \{eV\}; \{eV\}.
\type \verb|number|;\{\verb|number|,$\cdots$\};\{\verb|number|,$\cdots$\};\{\verb|number|,$\cdots$\} .
\example \verb|ldau_TLUJ=2;2,0,0;5,0,0;0,0,0|.
\syntax \verb|$aurl/?ldau_TLUJ|.
\end{itemize}

\item
\verb|natoms|
\begin{itemize}
\description Returns the number of atoms in the unit cell of the structure entry. The number can be non integer if partial occupation is considered within appropriate approximations.
\type\verb|number|.
\example \verb|natoms=12|.
\syntax \verb|$aurl/?natoms|.
\end{itemize}

\item
\verb|nbondxx|
\begin{itemize}
\description Nearest neighbors bond lengths of the relaxed structure per ordered set of species $A_i,A_{j\ge i}$.
For pure systems: $\left\{\rho\left[AA\right]\right\}$;
for binaries: $\left\{\rho\left[AA\right], \rho\left[AB\right], \rho\left[BB\right]\right\}$;
for ternaries: $\left\{\rho\left[AA\right], \rho\left[AB\right], \rho\left[AC\right], \rho\left[BB\right], \rho\left[BC\right], \rho\left[CC\right]\right\}$ and so on.
\type Set of $N_{species}(N_{species}+1)/2$ \verb|numbers|.
\units Natural units of the \verb|$code|, \Lunits.
\example \verb|nbondxx=1.2599,1.0911,1.0911,1.7818,1.2599,1.7818| (for
a three species entry).
\syntax \verb|$aurl/?nbondxx|. 
\end{itemize}

\item
\verb|nspecies|
\begin{itemize}
\description Returns the number of species in the system (e.g., binary = 2, ternary = 3, etc.).
\type \verb|number|.
\example \verb|nspecies=3|.
\syntax \verb|$aurl/?nspecies|.
\end{itemize}

\item
  \verb|Pearson_symbol_orig| (\verb|Pearson_symbol_relax|)
  \begin{itemize}
    \description Returns the Pearson symbol \cite{aflowBZ,Villars91} of the original-unrelaxed (relaxed) structure before (after) the calculation.
    \type \verb|string|.
    \example \verb|Pearson_symbol_orig=mS32| (\verb|Pearson_symbol_relax=mS32|).
    \syntax \verb|$aurl/?Pearson_symbol_orig| (\verb|$aurl/?Pearson_symbol_relax|).
    \tol {See entry} \verb|Bravais_lattice_orig| {or discussion about tolerances.}
  \end{itemize}
  
\item
\verb|positions_cartesian|
\begin{itemize}
\description Final Cartesian positions $(x_i,x_j,x_k)$ in the notation of the code.
\type Triplets (\verb|number|,\verb|number|,\verb|number|) separated by ``;'' for each atom in the unit cell.
\units Natural units of the \verb|$code|, e.g., in Cartesian coordinates (\AA) if the calculations were performed with {\small VASP} \cite{vasp}.
\example \verb|positions_cartesian=0,0,0;18.18438,0,2.85027;...|
\syntax \verb|$aurl/?positions_cartesian|. 
\end{itemize}

\item
\verb|positions_fractional|
\begin{itemize}
\description Final fractional positions $(x_i,x_j,x_k)$ with respect to the unit cell as specified in \verb|$geometry|.
\type Triplets (\verb|number|,\verb|number|,\verb|number|) separated by ``;'' for each atom in the unit cell.
\units adimensional
\example \verb|positions_fractional=0,0,0;0.25,0.25,0.25;...|
\syntax \verb|$aurl/?positions_fractional|. 
\end{itemize}

\item
\verb|pressure|
\begin{itemize}
\description Returns the external pressure selected for the simulation.
\type \verb|number|.
\units Natural units of the \verb|$code|, \Punits.
\example \verb|pressure=10.0|.
\syntax \verb|$aurl/?pressure|.
\end{itemize}

\item
\verb|prototype|
\begin{itemize}
\description Returns the AFLOW {\bf unrelaxed} prototype which was used for the calculation. The list can be accessed with the command ``{\sf aflow -{}-protos}'' or by consulting the online links.
The options are illustrated in the {\small AFLOW} manual \cite{aflowPAPER}.
Note that during the calculation, unstable structures can deform and lead to different relaxed configurations. It is thus {\bf imperative} for the user to make an elaborate analysis of the final structure to
pinpoint the right prototype to report. Differences in Bravais lattices, Pearson symbol, space groups, for the \verb|_orig| and \verb|_relax| versions are extremely useful for this task.
\type \verb|string|.
\example \verb|prototype=T0001.A2BC|.
\syntax \verb|$aurl/?prototype|.
\tol {See entry} \verb|Bravais_lattice_orig| {or discussion about tolerances.}
\end{itemize}

\item
\verb|PV_cell| (\verb|PV_atom|)
\begin{itemize}
\description Pressure multiplied by volume of the unit cell (of the atom).
\type \verb|number|.
\units Natural units of the \verb|$code|, \ENERGYunitatom.
\example \verb|PV_cell=12.13| (\verb|PV_atom=3.03|).
\syntax \verb|$aurl/?PV_cell| (\verb|$aurl/?PV_atom|).
\end{itemize}

\item
\verb|scintillation_attenuation_length|
\begin{itemize}
\description Returns the scintillation attenuation length of the compound in cm. See Refs. \cite{aflowSCINT,curtarolo:art46}.
\type\verb|real| number.
\example \verb|scintillation_attenuation_length=2.21895|.
\syntax \verb|$aurl/?scintillation_attenuation_length|.
\end{itemize}

\item
\verb|sg| (\verb|sg2|)
\begin{itemize}
\description Evolution of the space group of the compound \cite{tables_crystallography_A,bilbao}.
The first, second and third \verb|string| represent space group name/number before the first, after the first, and after
the last relaxation of the calculation.
\tol \verb|sg| values are calculated with 3.0\% and 0.5 deg tolerances for lengths and angles, respectively.
(\verb|sg2| is with 1.5\% and 0.25 deg).
Symmetry is cross validated through the internal engines of {\small AFLOW} \cite{aflowPAPER}, {\small PLATON} \cite{platon}, and {\small FINDSYM} \cite{findsym}.
\type Triplet \verb|string|,\verb|string|,\verb|string|.
\example \verb|sg=Fm-3m#225,Fm-3m#225,Fm-3m#225| (\verb|sg2=R-3c #167,R-3c #167,R-3c #167|).
\syntax \verb|$aurl/?sg| (\verb|$aurl/?sg2|).
\end{itemize}

\item
\verb|spacegroup_orig| (\verb|spacegroup_relax|)
\begin{itemize}
\description Returns the spacegroup number \cite{tables_crystallography_A} of the original-unrelaxed (relaxed) structure before (after) the calculation.
\tol Same as \verb|sg|.
\type \verb|number|.
\example \verb|spacegroup_orig=225| (\verb|spacegroup_relax=225|).
\syntax \verb|$aurl/?spacegroup_orig| (\verb|$aurl/?spacegroup_relax|).
\end{itemize}

\item
\verb|species|, \verb|species_pp|, \verb|species_pp_version|
\begin{itemize}
\description Species of the atoms, pseudopotentials species, and pseudopotential versions.
\type List of \verb|strings| separated by ``,''.
\example \verb|species=Y,Zn,Zr|, \verb|species_pp=Y_sv,Zn,Zr_sv|, \\
\verb|species_pp_version=Y_sv:PAW_PBE:06Sep2000,Zn:PAW_PBE:06Sep2000,Zr_sv:PAW_PBE:07Sep2000|.
\syntax \verb|$aurl/?species|, \verb|$aurl/?species_pp|, \verb|$aurl/?species_pp_version|.
\end{itemize}

\item
\verb|spin_cell| (\verb|spin_atom|)
\begin{itemize}
\description For spin polarized calculations, the total magnetization of the cell (magnetization per atom).
\type \verb|number|.
\units Natural units of the \verb|$code|, \Sunits.
\example \verb|spin_cell=2.16419| (\verb|spin_atom=0.541046|).
\syntax \verb|$aurl/?spin_cell| (\verb|$aurl/?spin_atom|).
\end{itemize}

\item
\verb|spinD|
\begin{itemize}
\description For spin polarized calculations, the spin decomposition over the atoms of the cell.
\type List of \verb|numbers| separated by ``,''.
\units Natural units of the \verb|$code|, \Sunits.
\example \verb|spinD=0.236,0.236,-0.023,1.005|.
\syntax \verb|$aurl/?spinD|. 
\end{itemize}

\item
\verb|spinD_magmom_orig|
\begin{itemize}
\description { For spin polarized calculations, string containing the values used to initialize the magnetic state for the {\it ab initio} calculation.}
\type {String containing the instruction passed to the {\it ab initio} code with spaces substituted by ``\_''.}
\units Natural units of the \verb|$code|.
\example \verb|spinD_magmom_orig=+5_-5_+5_-5|.
\syntax \verb|$aurl/?spinD_magmom_orig|. 
\end{itemize}

\item
\verb|spinF|
\begin{itemize}
\description For spin polarized calculations, the magnetization of the cell at the Fermi level.
\type \verb|number|.
\units Natural units of the \verb|$code|, \Sunits.
\example \verb|spinF=0.410879|.
\syntax \verb|$aurl/?spinF|. 
\end{itemize}

\item
\verb|stoichiometry|
\begin{itemize}
\description Similar to \verb|composition|, returns a comma delimited stoichiometry description of the structure entry in the calculated cell.
\type List of \verb|number| separated by ``,''.
\example \verb|stoichiometry=0.5,0.25,0.25|.
\syntax \verb|$aurl/?stoichiometry|.
\end{itemize}

\item
\verb|valence_cell_std| (\verb|valence_cell_iupac|)
\begin{itemize}
\description Returns standard valence (IUPAC valence, the maximum number of univalent atoms that may combine with the atoms \cite{iupac_gold}).
\type \verb|number|.
\example \verb|valence_cell_std=22| (\verb|valence_cell_iupac=12|)
\syntax \verb|$aurl/?valence_cell_std| (\verb|$aurl/?valence_cell_iupac|).
\end{itemize}

\item
\verb|volume_cell| (\verb|volume_atom|)
\begin{itemize}
\description Returns the volume of the unit cell (per atom in the unit cell).
\type \verb|number|.
\units Natural units of the \verb|$code|, \LLLunits.
\example \verb|volume_cell=100.984| (\verb|volume_atom=25.2461|).
\syntax \verb|$aurl/?volume_cell| (\verb|$aurl/?volume_atom|).
\end{itemize}

\end{itemize}

\end{widetext}

\section{Examples}
\label{examples}

\subsection{Generating a {free-energy zero temperature convex hull: OsTc}}
\label{example_phase_diagrams_2}

In this example we introduce the steps to generate a binary
{free-energy zero temperature convex hull} at zero temperature.
As an example, we choose the system OsTc \cite{monsterPGM,curtarolo:art70,curtarolo:art57}, and we illustrate the logical steps for obtaining it.
The user should prepare his/her own computer code to download and analyze the data as suggested.

{\bf 1.} Upon interrogation of the {\small AFLOWLIB}.org database ({\it database searches layer}, see Figures 1 and 2),
OsTc is found to be part of the {\it project-layer} with {\small AURL} \url{$aurl=aflowlib.duke.edu:AFLOWDATA/LIB2_RAW/}.
This \verb|$aurl| is translated into the {\small WEB} address \url{$web=http://aflowlib.duke.edu/AFLOWDATA/LIB2_RAW}.

{\bf 2.} The user downloads and parses the query
\url{$web/?keywords}. Being in a  {\it project-layer}, a better and faster alternative
is to download the entries' number and type with the queries \url{$web/?aflowlib_entries} and \url{$web/?aflowlib_entries_number}.
The user then parses \url{$web/?aflowlib_entries} and the string
\verb|Os_pvTc_pv| associated with the requested OsTc
{free-energy zero temperature convex hull}.

{\bf 3.} The user downloads and parses the \url{$web/Os_pvTc_pv/} part of a {\it set-layer}. The process can be accelerated by querying
\url{$web/Os_pvTc_pv/?aflowlib_entries} and \url{$web/Os_pvTc_pv/?aflowlib_entries_number} directly, to find further \verb|$aurl|.
The results are: \verb|$aflowlib_entries=1,2,3,4,..,657.AB,657.BA,..|,\\
\verb|$aflowlib_entries_number=260|. We enumerate and label these 260 entries with \verb|$entry|$_i$.

{\bf 4.} The user loops through \verb|$entry|$_i, \forall i$ and collects: \\
\url{$web/Os_pvTc_pv/}\verb|$entry|$_i$\url{/?stoichiometry}  and \\
\url{$web/Os_pvTc_pv/}\verb|$entry|$_i$\url{/?enthalpy_formation}.

{\bf 5}. Finally, the user collects the free-energies and plots the
convex hull as depicted in Figure 4.

{\bf 6}. The whole process can be performed with the {\small AFLOW} code.
The command ``{\sf aflow -{}-alloy OsTc -{}-update -{}-server=aflowlib.org}" connects to the appropriate server, downloads the information, calculates the
free-energy curve and prepares a PDF document with the appropriate information and hyperlinks to the individual entries.
See the {\small AFLOW} literature for more options
\cite{aflowPAPER}.
The user still has to double check the final
relaxed structure prototypes.
This is performed with a combination of: \\
\url{$web/Os_pvTc_pv/}\verb|$entry|$_i$\url{/?compound}, \\
\url{$web/Os_pvTc_pv/}\verb|$entry|$_i$\url{/?geometry}, \\
\url{$web/Os_pvTc_pv/}\verb|$entry|$_i$\url{/?positions_cartesian}, \\
\url{$web/Os_pvTc_pv/}\verb|$entry|$_i$\url{/?prototype}, \\
including files such as:\\
\url{$web/Os_pvTc_pv/}\verb|$entry|$_i$\url{/edata.orig.out}, and \\
\url{$web/Os_pvTc_pv/}\verb|$entry|$_i$\url{/edata.relax.out}, \\
and verifying the results by consulting appropriate prototype databases
(e.g., the {\it Naval Research Laboratory Crystal Structure database}, Ref. \cite{navy_crystal_prototypes}).

\begin{figure}[htb!]
 \centering
 \includegraphics[width=0.499\textwidth]{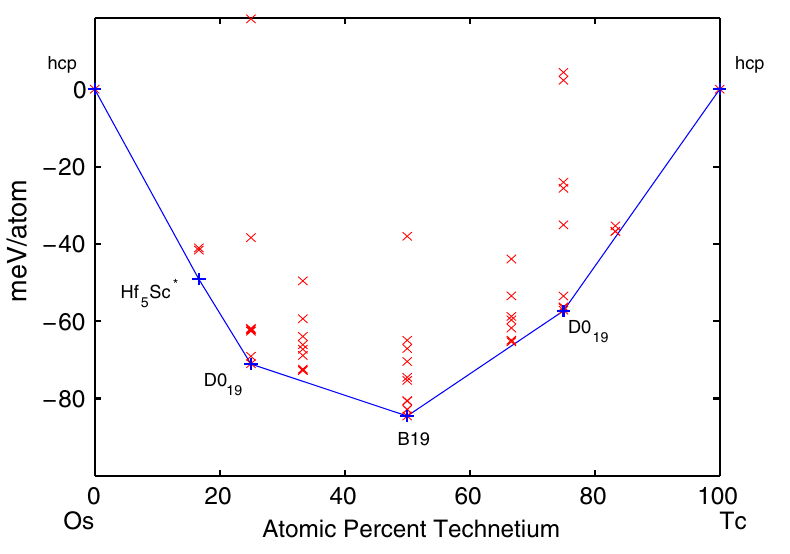}
 \vspace{-6mm}
 \caption{\small
  {Free-energy zero temperature convex hull} of OsTc automatically calculated through the
  {\small AFLOWLIB} {\small API} \cite{monsterPGM,curtarolo:art70,curtarolo:art57}.}
 \label{fig4}
\end{figure}

\subsection{{Generating a zero temperature phase-diagram of CoNbSi}}
\label{example_phase_diagrams_3}
{
In this example we introduce the steps to generate a ternary zero temperature phase diagram.
As an example, we choose the system CoNbSi \cite{curtarolo:art84}, and we illustrate the logical steps for obtaining it.
The user should prepare his/her own computer code to download and analyze the data as suggested.
}

{\bf 1.} Upon interrogation of the {\small AFLOWLIB}.org database ({\it database searches layer}, see Figures 1 and 2),
CoNbSi is found to be part of the {\it project-layer} with {\small AURL} \url{$aurl=aflowlib.duke.edu:AFLOWDATA/LIB3_RAW/}.
The stability of the ternary system also depends on the stability of three binary systems: CoNb, NbSi, and CoSi.
These are part of the {\it project-layer} with {\small AURL} \url{$aurl=aflowlib.duke.edu:AFLOWDATA/LIB2_RAW/}.
Ternary and binary \verb|$aurl|s are translated into the {\small WEB} address as \url{$web=http://aflowlib.duke.edu/AFLOWDATA/LIB3_RAW}, and 
\url{$web=http://aflowlib.duke.edu/AFLOWDATA/LIB2_RAW}, respectively.

\begin{figure}[htb!]
 \centering
 \includegraphics[width=0.49\textwidth]{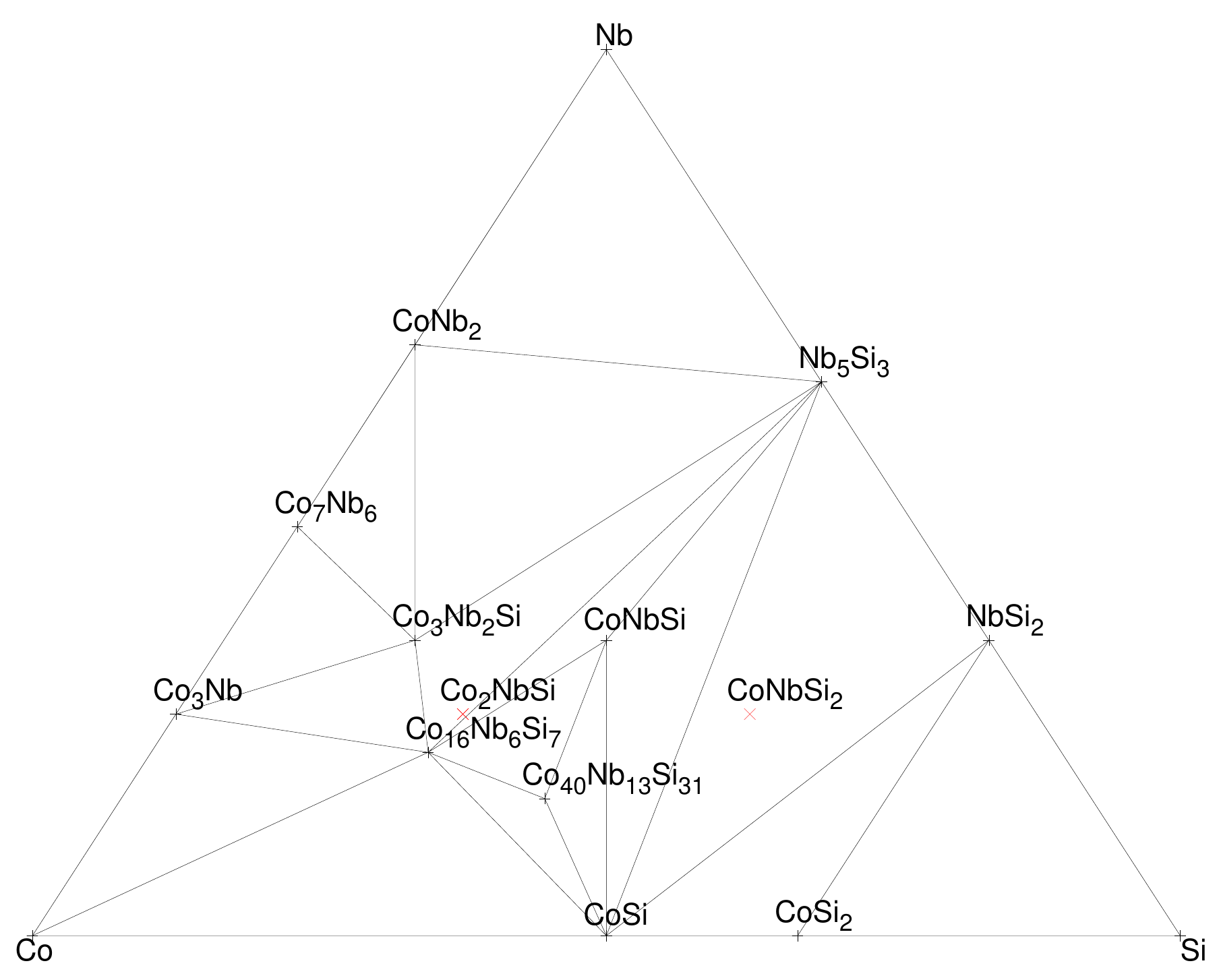}
 \vspace{-5mm}
 \caption{\small
  Zero-temperature phase diagram of CoNbSi automatically calculated through the {\small AFLOWLIB} {\small API}. Metastable half-Heuslers are marked with red stars \cite{curtarolo:art84}.}
 \label{fig5}
\end{figure}

{\bf 2.} Similarly to the previous example, the user downloads and parses the entries of \url{.../LIB3_RAW/CoNb_svSi/}, \url{.../LIB2_RAW/CoNb_sv/},
\url{.../LIB2_RAW/Nb_svSi/}, \url{.../LIB2_RAW/CoSi/}, which are part of the {\it set-layer}. 

{\bf 3.} The user loops through all the available entries and collects \url{/?stoichiometry} and \url{/?enthalpy_formation}, in
\url{.../LIB3_RAW/CoNb_svSi/}\verb|$entry|$_i$, \url{.../LIB2_RAW/CoNb_sv/}\verb|$entry|$_i$,
\url{.../LIB2_RAW/Nb_svSi/}\verb|$entry|$_i$, and \url{.../LIB2_RAW/CoSi/}\verb|$entry|$_i$.

{
{\bf 4}. The user calculates the convexity of the formation enthalpy landscape (we use {\small QHULL} \cite{qhull}) and plots the phase diagrams (we use {\small GNUPLOT} \cite{gnuplot}).
The diagram is depicted in Figure 5.
}

\subsection{Obtaining band structures}
\label{example_bands}
\begin{figure*}[t!]
 \centering
 \includegraphics[width=0.999\textwidth]{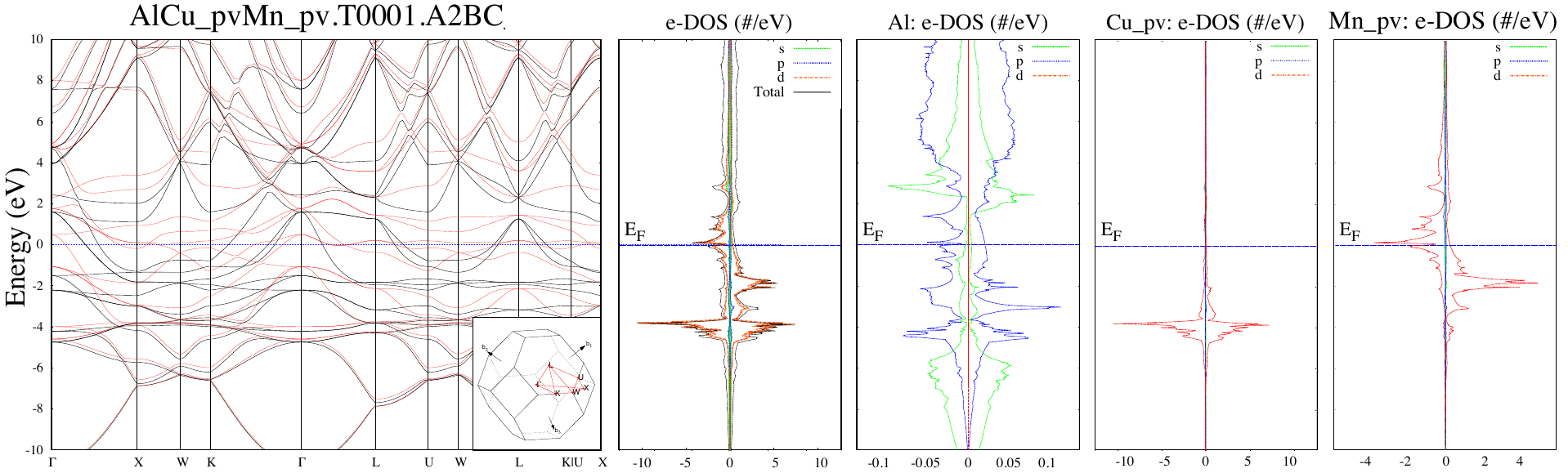}
 \vspace{-7mm}
 \caption{\small
  Band structure, total and partial electronic densities of states for Al$_\mathrm{2}$CuMn,
  as available through the {\small AFLOWLIB} {\small API}.
  The partial density of states is calculated only for inequivalent
  atomic positions [compound {\small AUID}=aflow:83cd2da0257e8def].}
 \label{fig6}
\end{figure*}

In this example, we introduce the steps to obtain the band structure and density of states plots for a calculated material.
As an example, we choose the compound Al$_\mathrm{2}$CuMn.
The user should prepare his/her own computer code to download and analyze the data as suggested.

{\bf 1.} Upon interrogation of the {\small AFLOWLIB}.org database ({\it database searches layer}, see Figures 1 and 2),
Al$_\mathrm{2}$CuMn is found to be part of the {\it project-layer} with {\small AURL} \url{$aurl=aflowlib.duke.edu:AFLOWDATA/LIB3_RAW/}.
This \verb|$aurl| is translated into the {\small WEB} address \url{$web=http://aflowlib.duke.edu/AFLOWDATA/LIB3_RAW}.

{\bf 2.} Within the {\it project-layer}, the user parses the query \url{$web/?aflowlib_entries}, which shows that the
string \verb|AlCu_pvMn_pv| is associated with the requested AlCuMn
ternary system.

{\bf 3.} The user parses \url{$web/AlCu_pvMn_pv/}. The query is part of a {\it set-layer}.
In this case, the set contains 10 {\it entries}, namely the calculation for the \verb|AlCu_pvMn_pv| system in
the prototypes \url|ICSD_57695.ABC|, \url{T0001.{A2BC,AB2C,ABC2}} (Heusler configurations),
\url{T0002.{A2BC,AB2C,ABC2}} (inverse Heusler), and \url{T0003.{ABC,BCA,CAB}} (half-Heusler).
The postfixes \url{A,B,C} indicate the positions of the species in the prototype; e.g. \url{A2BC} indicates the material Al$_\mathrm{2}$CuMn.

{\bf 4.}  Using the \url{?enthalpy_formation} and \url{?loop} queries for these 10 entries the user finds
which are stable and include a band structure calculation (indicated
by a negative formation enthalpy and the string \verb|bands| in the
\url{?loop} query output).
For this example, the user selects the {\it entry}
\url{$web/AlCu_pvMn_pv/T0001.A2BC/} for the compound
Al$_\mathrm{2}$CuMn, which satisfies both queries.

{\bf 5.} At the {\it calculation-layer}, the user finds the full \verb|aflowlib.out| entry.
By interrogating \url{$web/AlCu_pvMn_pv/T0001.A2BC/?files}, the user
obtains a list of all of the files available for download for this
calculation, including:
\begin{itemize}
\item
the input file for the calculation \url{$web/AlCu_pvMn_pv/T0001.A2BC/aflow.in},
\item
the spin polarized band structure with electronic density of states \url{$web/AlCu_pvMn_pv/T0001.A2BC/AlCu_pvMn_pv.T0001.A2BC.png}
\item
the Brillouin Zone in {\small AFLOW} notation \cite{aflowBZ} \url{AlCu_pvMn_pv.T0001.A2BC_BZ.png}
\item
the total electronic density of states \url{AlCu_pvMn_pv.T0001.A2BC_DOS.png}
\item
the partial electronic density of states for inequivalent positions
``Al"-\url{AlCu_pvMn_pv.T0001.A2BC_PEDOS_1_4_Al.png}, ``Cu"-\url{AlCu_pvMn_pv.T0001.A2BC_PEDOS_3_4_Cu.png}, and ``Mn"-\url{AlCu_pvMn_pv.T0001.A2BC_PEDOS_4_4_Mn.png}
\end{itemize}
A collage of these files is shown in Figure 6.

Also available in the {\it calculation-layer} for this entry are the
input and output files from the {\small VASP} and {\small AFLOW} runs,
that the user may extract from the output of the
\url{$web/AlCu_pvMn_pv/T0001.A2BC/?files} command.

\subsection{Synergy of experimental and calculation data on a rare prototype}
\label{synergyexample}

The experimental data on binary alloys contains many gaps. It also
presents a huge panoply of structural prototypes, ranging from very
common ones, appearing in hundreds of compounds, to very rare ones
appearing in just a few systems. HT
calculations can be used to bridge those gaps and provide a more
complete picture about the existence of yet unobserved compounds and their
structures. They can also considerably extend the predicted range of
those rare prototypes, indicating their existence in a larger set of
binary systems. One such example studied the prevalence of the Pt$_8$Ti
prototype. This structure has been experimentally observed in 11
systems, but a high-throughput search over all of the binary transition
intermetallics revealed it should be stable at low temperatures in 59
systems \cite{curtarolo:art56}. The study verified all the experimental occurrences while
offering additional predictions, including a few surprising ones in
supposedly well-characterized systems (e.g., Cu-Zn). This example serves as a striking demonstration
of the power of the high-throughput approach. In this section we
present a new example, discussing recent reports observing the rare
prototype Pd$_4$Pu$_3$ in a few transition metal binaries and
computationally predicting a considerable extension of its stability or
metastability in such systems.

\begin{figure}[h!]
\includegraphics[width=0.45\textwidth]{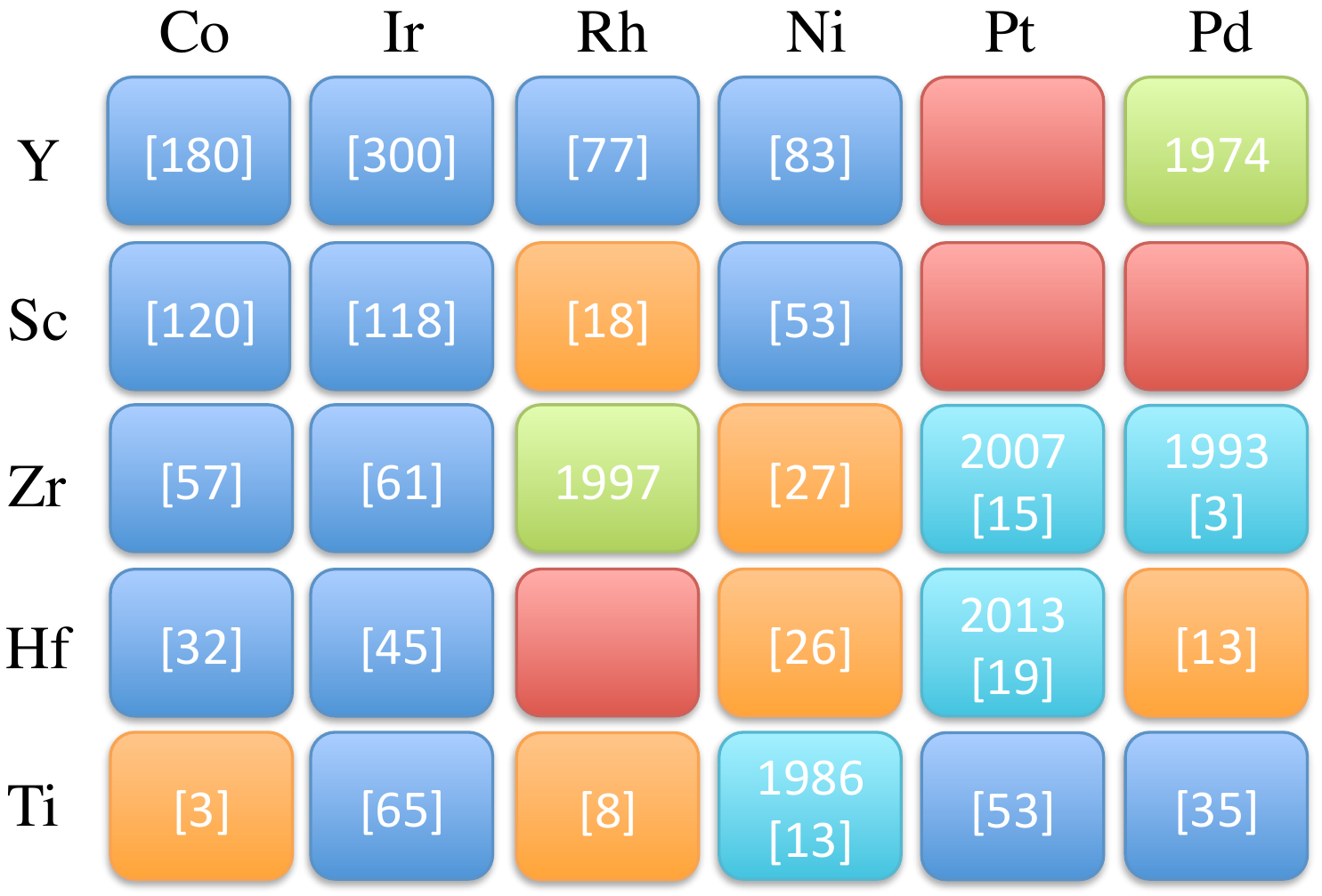}
\caption{\small Pettifor-type structure map of the Pd$_4$Pu$_3$ phase in transition metal binary systems.
  Both axes are labeled by increasing Mendeleev number after Pettifor \cite{pettifor:1986}.
  Colors denote reported compounds with indicated year of discovery (green and light blue) and prediction of unreported compounds (red and
  orange) found to be stable (green and red) or metastable (light blue and orange) in the calculations. Systems where the structure is unstable
  (formation enthalpy of more than 30 meV/atom above the convex hull) are denoted in blue.
  The square parentheses denote the formation enthalpy of metastable and unstable structures above the convex hull of the respective system.}
\label{fig7}
\end{figure}

The Pd$_4$Pu$_3$ (hR14, space group \#148) was first observed in its eponymous system in 1967 \cite{Kutaitsev-AE1967,Cromer-ActaCrystB1973}.
It has since been reported in 37 additional binary systems, mostly of a lanthanide or an actinide with the elements Pt or Pd. \cite{Villars2013}.
Only 6 compounds of this prototype have been reported in transition metal binary systems: Ni$_4$Ti$_3$
\cite{Saburi-JLCM1986,Tirry-ActaCrystB2006}, Pd$_4$Y$_3$ \cite{Palenzona-JLCM1974}, Pd$_4$Zr$_3$
\cite{Bendersky-JAC1993}, Pt$_4$Zr$_3$ \cite{Stalick-JAC2007}, Rh$_4$Zr$_3$ \cite{Bendersky-JAC1997}, and
most recently Hf$_3$Pt$_4$ \cite{Stalick_HfPt_2013}.
In these compounds, one component is a 3B or 4B element and the other is from the ninth
or tenth column of the periodic table. In this example we wish to
examine the possible appearance of this prototype in all transition metal binary systems
of these columns (30 systems). This can be done in a few steps,
as follows:

{\bf 1.} Consulting the complete list of structure designations of {\small
  AFLOW} with the command ``{\sf aflow -{}-protos}'' or by the online
links, the user finds the label of the prototype, in this case \verb|655.AB|
or \verb|655.BA| (depending on the order of the species).

{\bf 2.} Using
\url{$aurl=aflowlib.duke.edu:AFLOWDATA/LIB2_RAW/?aflowlib_entries}
the user finds the entry name
for each of those 30 systems in the {\it set-layer}. Then, using
\url{$aurl=aflowlib.duke.edu:AFLOWDATA/LIB2_RAW/XXX/?aflowlib_entries}
for each of those names the user finds the calculations of
the desired prototype in the {\it calculaiton-layer} (indicated
by the string \verb|655.BA| or \verb|655.BA| in the query output).

{\bf 3.} Following the steps of example \ref{example_phase_diagrams_2}
the user constructs the convex hull for each of these systems and
finds the position of the desired structure in it, as either stable, metastable or
unstable.

Following these steps for the Pd$_4$Pu$_3$ prototype we find that it
appears as a low temperature stable
compound in six systems, two
reported in experiments and four newly predicted ones. The
structure is also found to be metastable (with less than 30meV/atom above
the respective system convex hull) in ten
systems, of which it was reported in four by experiments,  and is predicted in six
additional ones. Among the predicted phases, three compounds of the same stoichiometry,
Pt$_4$Y$_3$, Hf$_3$Pd$_4$ and Hf$_3$Rh$_4$, are reported with an
unknown structure in the experimental literature but identified with
the Pd$_4$Pu$_3$ structure in the calculations.
Overall, the calculation extends the prevalence of this prototype
(stable or metastable) among transition metal binaries from five
systems to sixteen.
Figure \ref{fig7} summarizes these results.

\begin{widetext}
\subsection{Bash api.sh example}
\label{example_bash}
This ``bash" script example \verb|api.sh| downloads an \verb|aflowlib.out| entry for the {\it project-}, {\it set-}, or {\it calculation-layers} of the binary alloys.
\begin{verbatim}
#!/bin/bash                                                             # run sh ./api.sh
SERVER='http://aflowlib.duke.edu'                                       # server name
PROJECT='AFLOWDATA/LIB2_RAW/'                                           # project name
#URL=$SERVER'/'$PROJECT                                                 # project-layer
URL=$SERVER'/'$PROJECT'Os_pvTc_pv/'                                     # set-layer
#URL=$SERVER'/'$PROJECT'/Os_pvTc_pv/657.AB/'                            # calculation-layer
IFS=',';                                                                # options
for key in $(wget -q -O - ${URL}?keywords);                             # get all keywords
do                                                                      # loop keywords
    val=$(wget -q -O - ${URL}?${key});                                  # assign one keyword
    echo "${key}=${val}";                                               # print keyword
done                                                                    # end loop
\end{verbatim}
\subsection{Python api.py example}
\label{example_python}
This ``python3" script example \verb|api.py|
downloads an \verb|aflowlib.out| entry for the {\it project-}, {\it set-}, or {\it calculation-layers} of the Heusler alloys database.
\begin{verbatim}
#!/usr/bin/python3                                                      # python3
import json                                                             # preamble 
from urllib.request import urlopen                                      # preamble
SERVER='http://aflowlib.duke.edu'                                       # server name 
PROJECT='AFLOWDATA/LIB3_RAW/'                                           # project name
#URL=SERVER+'/'+PROJECT                                                 # project-layer
#URL=SERVER+'/'+PROJECT+'AlCu_pvMn_pv/'                                 # set-layer
URL=SERVER+'/'+PROJECT+'AlCu_pvMn_pv/T0001.A2BC/'                       # calculation-layer
entry=json.loads(urlopen(URL+'?format=json').readall().decode('utf-8')) # load
for key in entry:                                                       # loop keys
    print( "{}={}".format(key, entry[key]) )                            # print key
\end{verbatim}

\end{widetext}

\section{Updates beyond version 1.0}
\label{updates}
Standards, like databases, are only as good as the updates they receive when new quantities and descriptors become available.
The list of \verb|keywords| available in the current version of the standard is far from being complete for rational materials design. 
The user is invited to search and consult appropriate {\small API} specifications {\it addenda}, which will be 
published periodically through the consortium website {\small AFLOWLIB}.org.
The entries' {\small API} version can be found by inquiring the keyword \verb|$aurl/?data_api| as described above. 

\section{Conclusion}
\label{conclusions}

The {\small AFLOWLIB} {\small API} provides a simple and powerful tool for accessing a large set of simulated materials properties data.
This will allow the community to make use of  {\small AFLOWLIB} to the fullest
extent possible, through search formats allowing complete accessibility of
the database contents at all levels and integration of search results
into externally formulated workflows. Such workflows may execute any
type of investigation on the obtained data, ranging from a simple
study of the properties of a specific material to extensive
statistical analyses of whole structure classes for materials prediction.
The full provenance of the data produced is provided, following a standard of reproducible and transparent
scientific data sharing, to facilitate its straightforward reproduction
and extension.

The {\small AFLOWLIB} database is growing continually by updating existent alloy libraries and adding new ones (e.g., recent
attention is focused on ternary systems and electronic properties).
The new {\small API} described in this paper is built on top of the {\small AFLOW}
framework, developed to create the database and to interrogate it, but it can be easily extended to other materials design environments.
It is constructed as a federatable tool to maximize the utility of the database
to the scientific community and expedite scientific collaboration with particular emphasis on reproducibility, accessibility and attribution.

\section{Acknowledgments}
\label{acknowledgements}
The authors thank Dr. K. Yang, Dr. J. Warren, Dr. D. Irving, Dr. G. Hart, Dr. S. Sanvito, Dr. L. Kronik, Dr. A. Kolmogorov, Dr. M. Mehl,  Dr. N. Mingo, Dr. J. Carrete, Dr. A. Natan, 
Dr. M. Fornari, Dr. O. Isayev, Dr. A. Tropsha, {Dr. K. Persson, Dr. G. Ceder}, and Dr. A. Stelling for useful comments.
This work is partially supported by DOD-ONR (N00014-13-1-0635, N00014-11-1-0136, N00014-09-1-0921), 
NIST \#70NANB12H163 and by the Duke University---Center for Materials Genomics.
C.T. and S.C. acknowledge partial support by DOE (DE-AC02-05CH11231), specifically the BES program under Grant \#EDCBEE.
The consortium {\small AFLOWLIB}.org acknowledges the Fulton Supercomputer Center and the CRAY corporation for computational assistance.


\end{document}